\documentclass[aps,prd,onecolumn,superscriptaddress,nofootinbib]{revtex4-2}

\usepackage{amsmath,amssymb,bm,mathtools}
\usepackage{booktabs}
\usepackage{graphicx}
\usepackage{hyperref}
\hypersetup{
  colorlinks=true,
  linkcolor=blue,
  citecolor=blue,
  urlcolor=blue,
  pdfstartview={XYZ null null 1}
}

\begin{document}

\title{Constraining the gravitational-wave breathing mode\\
with LVK antenna patterns and GWTC data}

\author{Hong-Bo Jin}
\email{hbjin@bao.ac.cn}
\affiliation{National Astronomical Observatories, Chinese Academy of Sciences, Beijing 100101, China}
\affiliation{The International Centre for Theoretical Physics Asia-Pacific, University of Chinese Academy of Sciences, Beijing 100190, China}

\author{Yanyan Zheng}
\email{zytfc@mst.edu}
\affiliation{Missouri University of Science and Technology, Rolla, Missouri 65409, USA}

\date{\today}

\begin{abstract}
Metric theories of gravity admit a massless breathing scalar $b$ besides the tensor modes $+$ and $\times$. Many scalar--tensor models predict a much smaller breathing amplitude; gravitational quantum field theory (GQFT) instead allows $h_b$ comparable to the tensor amplitudes. We evaluate the LIGO--Virgo--KAGRA (LVK) antenna patterns with PyCBC, including Michelson suppression of $b$. Here $F_b$ is the breathing antenna pattern and $\mathcal{S}_b=\sum_d F_b^{d\,2}$ is the network breathing power. At the H1 zenith at the GW150914 epoch, $F_b\sim 5\times 10^{-6}$; for $72$ Gravitational-wave Transient Catalog (GWTC) events with packaged strain the median inclination gain in $\mathcal{S}_b$ is $\sim 34\times$. Ranking multi-detector GWTC events (network $\mathrm{SNR}\ge 8$, $\ge 2$ sites) by $\mathrm{SNR}\sqrt{\mathcal{S}_b^{\,*}}$ identifies loud O4 triggers and GW170817 as geometric targets. Parameter-estimation (PE) weighted $\mathcal{S}_b$ on twelve events shows that the true sky can suppress this ranking (GW170817) and that GW200129\_065458, GW150914, and GW190814 have the highest PE-weighted $\mathcal{S}_b$ for $\xi$ in $h_b=\xi h_+$ (GR: $\xi\equiv 0$). An HLV design-noise injection grid gives $\ln B(\mathrm{GR}{+}b/\mathrm{GR})\approx 13$ ($4$) for $\xi=1$ ($0.5$) on an inclined sky ($\mathcal{S}_b\sim 0.18$); near zenith ($\mathcal{S}_b\sim 0.016$) the $\xi$ posterior remains unconstrained. The GWTC limits are semi-joint: masses and spins fixed at PE medians, while $d_L$, $\theta_{JN}$, sky, and $\xi$ vary within PE-informed priors. Eleven of twelve events are analyzed (GW170817 omitted: PE sky on the breathing null). None favors $\mathrm{GR}{+}b$; $|\xi|_{90\%}\approx 0.15$, $0.18$, and $0.23$ for GW150914, GW200129\_065458, and GW200311\_115853 disfavor $|\xi|\sim\mathcal{O}(1)$, while weaker bounds (including GW190814) still allow that scale.
\end{abstract}

\maketitle

\section{Introduction}

Gravitational-wave (GW) polarization distinguishes metric theories of gravity. For a weak plane null wave, at most six polarizations are admitted~\cite{Eardley1973}: the tensor modes $+$ and $\times$, the vector modes $x$ and $y$, and the scalar modes (breathing $b$ and longitudinal $l$). General relativity (GR) allows only $+,\times$, which have been identified in LIGO~\cite{LIGO}, Virgo~\cite{Virgo} and KAGRA~\cite{KAGRA:2022qtq} (LVK) observations~\cite{Abbott2019TGR,Abbott2021TGR,GWTC4TGR1,GWTC4TGR2,GWTC4TGR3,GWTC5TGR}. Scalar--tensor theories such as Brans--Dicke can generate an additional breathing mode~\cite{Brans1961,Brans1962}, but solar-system tests tightly constrain the scalar coupling (Cassini implies $\omega_{\mathrm{BD}}\gtrsim 4\times 10^{4}$)~\cite{Bertotti2003,Will2014}, so that for observationally viable parameters the predicted breathing amplitude is typically much smaller than the tensor sector~\cite{Will2014,Berti2015}. For binary black holes the scalar dipole is further suppressed when the companions have (nearly) equal gravitational sensitivities~\cite{Berti2015}. The relative breathing amplitude $\xi$ is defined for compact binary coalescence (CBC) templates by $h_b=\xi\,h_+$; GR corresponds to $\xi\equiv 0$. A nondetection of $b$ in present data is therefore expected when $|\xi|\ll 1$, even before the geometric Michelson suppression of the breathing antenna pattern $F_b$ (Sec.~\ref{sec:ground}). Gravitational quantum field theory (GQFT)~\cite{Wu2023,Gao2024,Gao:2025aye,XuJinWu2025} admits a massless breathing mode that can remain comparable to the tensor sector, which we take as a phenomenological scale $|\xi|\sim\mathcal{O}(1)$. The amplitude $\xi$ is constrained with Gravitational-wave Transient Catalog (GWTC) data; GQFT is used only as that scale.

The six-mode antenna-pattern formalism for interferometers, including the scalar breathing response, is developed in Ref.~\cite{Nishizawa:2009bf} and is used here when evaluating the LVK antenna patterns with the PyCBC toolkit~\cite{pycbc_software,Usman2016}. Network separability of tensor, vector, and scalar modes for compact binaries is discussed in Ref.~\cite{Takeda2018}. Null-stream methods isolate non-tensor content for networks with known sky position~\cite{Hagihara2018}, with upper bounds from GW170817~\cite{Hagihara2019}. Bayesian tests for mixtures of tensor and alternative polarizations via null streams are given in Ref.~\cite{Pang2020}. Cataloged LVK tests of GR with GWTC-1 and GWTC-3~\cite{Abbott2019TGR,Abbott2021TGR} compare pure-tensor, pure-scalar, and pure-vector hypotheses and find no evidence for non-GR polarizations; those analyses leave unconstrained the mixed amplitude $\xi$ of a breathing component tied to $h_+$. A Gaussian-process null-stream search of GWTC-3 three-detector events sets model-agnostic upper limits on fractional nonstandard strain (as low as $\sim 0.28$ at $90\%$ credibility for selected events)~\cite{Cheung2025}, and uses a parameterization other than $h_b=\xi h_+$. Polarization tests with the stochastic background~\cite{Callister2017} apply to a different source population. Breathing versus tensor responses for LISA-/Taiji-like space detectors have been mapped at $|\xi|\sim\mathcal{O}(1)$~\cite{XuJinWu2025}. The present analysis is the ground-based counterpart: Michelson differential readout suppresses $F_b$ near zenith. The network breathing power is $\mathcal{S}_b=\sum_d F_b^{d\,2}$. The amplitude $\xi$ is constrained with PyCBC antenna patterns, $\mathcal{S}_b$ on parameter-estimation (PE) skies, an injection grid in $(\xi,\mathcal{S}_b)$, and GWTC strain. The GWTC limits are \emph{semi-joint}: masses and spins are fixed at preferred PE medians, while the luminosity distance $d_L$, inclination $\theta_{JN}$, sky location, and $\xi$ are sampled within PE-informed priors, so that $\mathcal{S}_b$ enters the inference with the remaining CBC parameters held at those PE medians.

When the wave vector is nearly orthogonal to the detector plane---where the tensor response is strongest---differential readout of an L-shaped Michelson interferometer nearly cancels the isotropic $b$ mode~\cite{Nishizawa:2009bf,Takeda2018}. Below we review the LVK antenna-pattern formulas, evaluate them with PyCBC, quantify $\mathcal{S}_b$ on GWTC skies, and compare nested models GR versus GR+$b$ for injections and selected high signal-to-noise ratio (SNR) events under the semi-joint setup above. Section~\ref{sec:pol} reviews the polarization basis. Section~\ref{sec:ground} derives the L-shaped detector response. Section~\ref{sec:pycbc} matches the formulas to PyCBC. Section~\ref{sec:bmode} discusses observability of the $b$ mode. Section~\ref{sec:network} outlines the analysis strategy. Section~\ref{sec:results} reports geometric scans, PE-weighted $\mathcal{S}_b$, and semi-joint constraints on $\xi$. Section~\ref{sec:concl} summarizes.

\section{Polarization decomposition and geodesic deviation}
\label{sec:pol}

A weak plane null wave acts on test masses through the geodesic-deviation equation, with six independent components of $R_{i0j0}$ that map onto the metric polarizations $\{+,\times,x,y,b,l\}$~\cite{Eardley1973,Hyun2019}. We retain the tidally relevant set $\{+,\times,b\}$, set $h_l=0$, and omit vector modes from the signal model. The breathing amplitude is the phenomenological $\xi$ with $h_b=\xi h_+$ introduced above.

A far-zone plane wave is written as~\cite{Nishizawa:2009bf}
\begin{equation}
  \mathbf{h}(t,\mathbf{x})
  =\sum_A h_A\!\left(t-\hat{k}\cdot\mathbf{x}/c\right)\varepsilon_A(\hat{k}),
  \label{eq:h-pol}
\end{equation}
where $A$ labels polarizations, $c$ is the speed of light, and the polarization tensors are built from the right-handed wave triad $\{\hat{k},\hat{m},\hat{n}\}$:
\begin{align}
  \varepsilon_+ &= \hat{m}\otimes\hat{m}-\hat{n}\otimes\hat{n},\quad
  \varepsilon_\times = \hat{m}\otimes\hat{n}+\hat{n}\otimes\hat{m},\nonumber\\
  \varepsilon_b &= \hat{m}\otimes\hat{m}+\hat{n}\otimes\hat{n}.
  \label{eq:eps}
\end{align}
(The vector and longitudinal tensors are omitted from the signal model.) Equatorial right ascension $\alpha$, declination $\delta$, and polarization angle $\psi$ fix the triad in the LIGO Algorithm Library (LAL)/PyCBC convention,
\begin{align}
  \hat{k} &= (\cos\delta\cos\alpha,\cos\delta\sin\alpha,-\sin\delta),\nonumber\\
  \hat{u} &= (\sin\delta\cos\alpha,\sin\delta\sin\alpha,\cos\delta),\nonumber\\
  \hat{v} &= (-\sin\alpha,\cos\alpha,0),\nonumber\\
  \hat{m} &= \cos\psi\,\hat{u}+\sin\psi\,\hat{v},\quad
  \hat{n} = -\sin\psi\,\hat{u}+\cos\psi\,\hat{v},
  \label{eq:sky}
\end{align}
with $\hat{k}$ pointing from the source toward the observer and $\psi=0$ aligning $\hat{m}$ with $\hat{u}$.

\section{Response of terrestrial L-shaped interferometers}
\label{sec:ground}

For unit arm vectors $\hat{e}_X,\hat{e}_Y$, the Michelson detector tensor is
\begin{equation}
  D_{ij}
  =\frac{1}{2}\Bigl(\hat{e}_X{}_i\hat{e}_X{}_j - \hat{e}_Y{}_i\hat{e}_Y{}_j\Bigr),
  \label{eq:D}
\end{equation}
and in the long-wavelength limit the strain is
\begin{equation}
  h(t)=\sum_A F_A(\alpha,\delta,\psi,t)\, h_A(t),
  \qquad
  F_A = D^{ij}\,(\varepsilon_A)_{ij},
  \label{eq:hdet}
\end{equation}
or, including vector and scalar modes~\cite{Nishizawa:2009bf},
\begin{equation}
  h(t)
  = F_+ h_+ + F_\times h_\times
  + F_x h_x + F_y h_y
  + F_b h_b + F_l h_l.
  \label{eq:h-all}
\end{equation}
The breathing tensor $\varepsilon_b$ is isotropic in the wavefront plane. At zenith, with equal orthogonal arms, the arm unit vectors projected onto the wavefront plane, $\hat{e}_X^\perp$ and $\hat{e}_Y^\perp$, are equal in length, so
\begin{equation}
  F_b
  = D^{ij}(\varepsilon_b)_{ij}
  =\frac{1}{2}\Bigl(|\hat{e}_X^\perp|^2 - |\hat{e}_Y^\perp|^2\Bigr)=0,
\end{equation}
while $|F_+|$ and $|F_\times|$ are typically large: maximal tensor sensitivity coincides with a breathing null. For arbitrary sky location,
\begin{equation}
  F_b(\alpha,\delta,\psi)
  =\frac{1}{2}\Bigl[
    (\hat{e}_X\cdot\hat{m})^2+(\hat{e}_X\cdot\hat{n})^2
    -(\hat{e}_Y\cdot\hat{m})^2-(\hat{e}_Y\cdot\hat{n})^2
  \Bigr],
\end{equation}
so $F_b$ measures only the difference of the arm projections. A single-arm breathing strain can remain nonzero; the differential arm-length combination $\delta L_X-\delta L_Y$ suppresses it. Terrestrial LVK products are predominantly differential, so the network geometry enters the inference.

As Earth rotates, $F_A=F_A(\mathrm{GMST}(t)-\alpha,\delta,\psi)$, where $\mathrm{GMST}$ is Greenwich mean sidereal time. For the short CBC signals analyzed below, $F_A$ is held fixed over each analysis window.

\section{Correspondence with the PyCBC analysis toolkit}
\label{sec:pycbc}

Antenna patterns for H1, L1, V1, and K1 are evaluated with PyCBC~\cite{pycbc_software,Usman2016}, which implements the contractions of Eqs.~\eqref{eq:D}--\eqref{eq:h-all} (including geocentric time delays via LALSimulation in the LAL stack~\cite{lalsuite}). With wave-basis vectors $(\hat{x},\hat{y},\hat{z})$ corresponding to $(\hat{m},\hat{n},\hat{k})$,
\begin{align}
  F_+ &= x_i D^{ij} x_j - y_i D^{ij} y_j, \quad
  F_\times = x_i D^{ij} y_j + y_i D^{ij} x_j, \nonumber\\
  F_b &= x_i D^{ij} x_j + y_i D^{ij} y_j,
  \label{eq:pycbc-F}
\end{align}
Extra-polarization analyses therefore reuse these $F_A$: the geometric response is fixed by the detectors and sky, while theory dependence enters through $\{h_A\}$. All sky maps and $\mathcal{S}_b$ evaluations below use this PyCBC realization.

\section{Observability of the breathing mode in the LVK network}
\label{sec:bmode}

\subsection{Relative breathing amplitude $\xi$}

Tensor polarizations $(h_+,h_\times)$ for compact binaries are obtained from standard CBC waveform families in the LAL/PyCBC stack (phenomenological and related models used for the injections and semi-joint analyses below). Throughout Sec.~\ref{sec:results} the breathing mode is parameterized as
\begin{equation}
  h_b(t)=\xi\,h_+(t),
  \label{eq:xi}
\end{equation}
so that $h_b$ inherits the plus-mode inclination dependence and phase from the CBC template. GR corresponds to $\xi\equiv 0$; $|\xi|\sim\mathcal{O}(1)$ denotes a breathing strain comparable to $h_+$, the phenomenological scale taken from GQFT's scalar sector. Scalar radiation patterns are theory dependent; setting $h_b=\xi h_+$ keeps $\xi$ aligned with the GWTC templates.

\subsection{Single-site geometry and network $\mathcal{S}_b$}

For a single L-shaped interferometer, define the tensor and breathing antenna powers
\begin{equation}
  \mathcal{S}_T = F_+^2 + F_\times^2,\qquad
  \mathcal{S}_b = F_b^2.
\end{equation}
Near zenith, $\mathcal{S}_T$ is large while $\mathcal{S}_b\approx 0$; inclined skies restore $F_b$ at the cost of reduced $\mathcal{S}_T$. With three or more detectors of distinct orientation,
\begin{equation}
  \mathbf{h}_{\mathrm{net}}(t)
  =\mathbf{F}\, \mathbf{h}_{\mathrm{pol}}(t)+\mathbf{n}(t),
\end{equation}
where $\mathbf{n}(t)$ is the detector noise and the $d$th row of the network response matrix $\mathbf{F}$ is $(F_+^d,F_\times^d,F_b^d,\ldots)$. When $\mathrm{rank}(\mathbf{F})\ge 3$, nested hypotheses $\mathcal{H}_{\mathrm{GR}}$ ($\xi\equiv 0$) versus $\mathcal{H}_{\mathrm{GR}+b}$ ($\xi$ free) can be compared via Bayes factors. The H1--L1 baseline and the arm orientations of V1/K1 remove the single-detector zenith null for generic skies; which detectors operate at a given epoch then set the \emph{network} breathing power $\mathcal{S}_b=\sum_d F_b^{d\,2}$ (the same symbol as the single-site $F_b^2$, hereafter the summed network quantity unless a single site is specified).

\section{LVK network analysis strategy}
\label{sec:network}

The analysis injects or models $h_b=\xi h_+$ on top of a GR waveform family, projects onto each detector with the corresponding antenna patterns and geocentric time delays, and compares nested hypotheses $\mathcal{H}_{\mathrm{GR}}$ versus $\mathcal{H}_{\mathrm{GR}+b}$ with bilby.

In the scalar sector we set $h_l=0$ and retain only $F_b$. For the short CBC windows used below, $F_A$ is constant in time. When Eq.~\eqref{eq:h-pol} holds, waveform-model dependence enters mainly through $\{h_A\}$ rather than through $F_A$.

\section{Geometric and Bayesian results for the LVK network}
\label{sec:results}
The antenna-pattern formulas in Sec.~\ref{sec:pycbc} are evaluated for the LVK interferometers using PyCBC and applied to the GWTC metadata~\cite{Lgwtc1,gwtc2,gwtc3,LIGOScientific:2025gwtc4,LIGOScientific:2026gwtc5}, downloaded from the Gravitational Wave Open Science Center (GWOSC)~\cite{LVK_O4b_OpenData_2026}. We select GWTC events that are observed by at least two detectors, have a network $\mathrm{SNR}\ge 8$, and have packaged $4\,\mathrm{kHz}$ event-page strain ($32\,\mathrm{s}$ files), yielding a total of $72$ events.

The geometric scan and PE-weighted $\mathcal{S}_b$ below quantify the terrestrial breathing response that enters the subsequent Bayesian analyses of $\xi$.

\subsection{Zenith null of $F_b$}
\begin{figure}[t]
\includegraphics[width=\columnwidth]{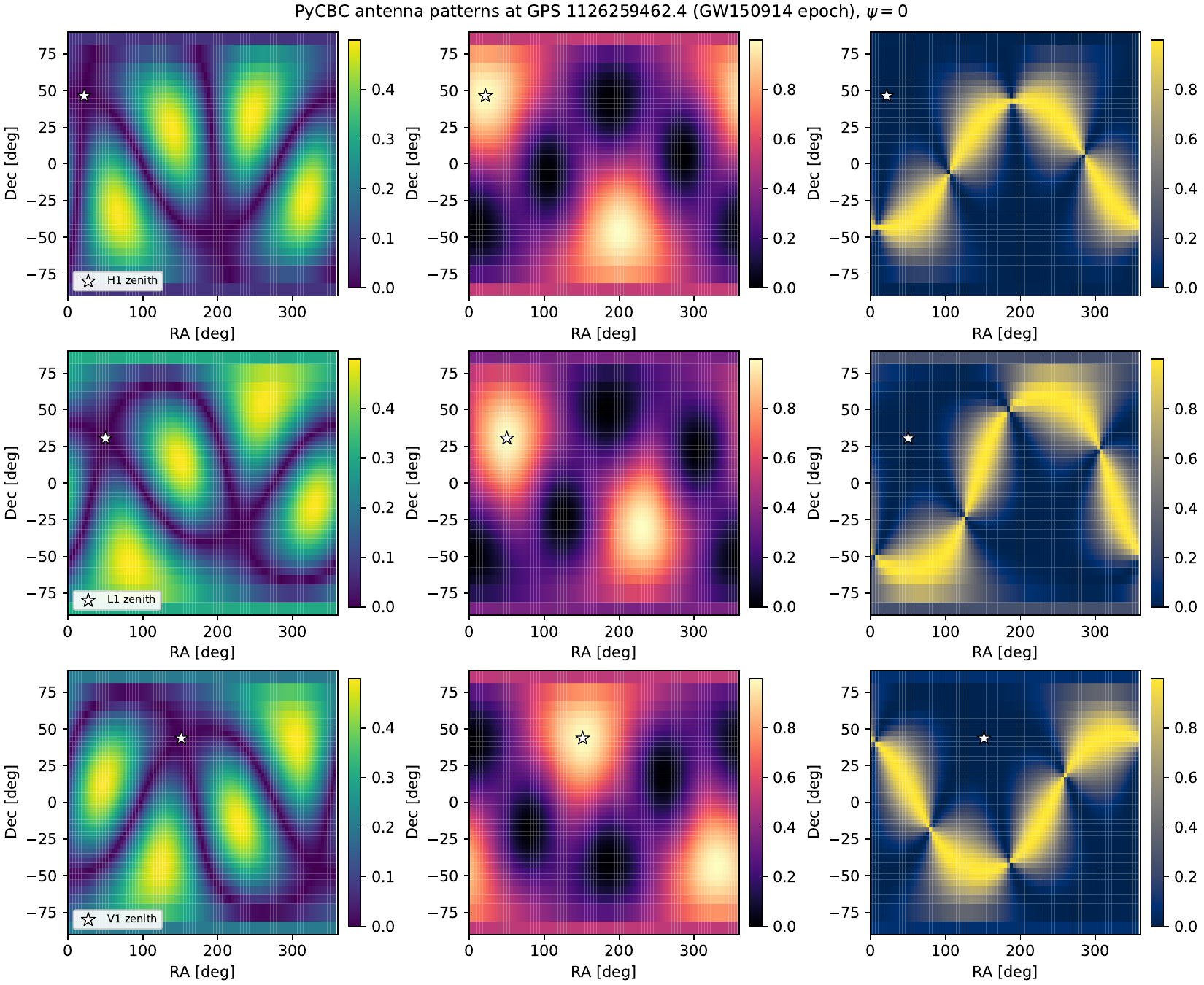}
\caption{\label{fig:H1}
Antenna patterns of LIGO Hanford (H1, top row), Livingston (L1, middle
row), and Virgo (V1, bottom row) at the GW150914 epoch, evaluated with
PyCBC (GPS $1126259462.4$, $\psi=0$) in equatorial coordinates (RA, Dec).
Left to right: $|F_b|$; tensor power $\mathcal{S}_T=F_+^2+F_\times^2$;
ratio $\mathcal{S}_b/\mathcal{S}_T$ with $\mathcal{S}_b=F_b^2$.
Stars mark each detector's local zenith (H1: RA $\simeq 21.3^\circ$, Dec
$\simeq 46.5^\circ$; L1: RA $\simeq 50.0^\circ$, Dec $\simeq 30.6^\circ$;
V1: RA $\simeq 151.3^\circ$, Dec $\simeq 43.6^\circ$),
which in all three cases lies in the large-$\mathcal{S}_T$, near-null-$F_b$
sector. The suppression appears at each site, rotated to the sky position
set by the detector's location and arm orientation.}
\end{figure}

At the local zenith of H1 and the GW150914 epoch one finds
\begin{equation}
  F_b \simeq 5\times 10^{-6},\qquad
  |F_+|\simeq 0.30,\qquad
  |F_\times|\simeq 0.95,
\end{equation}
in agreement with the analytic differential cancellation of an isotropic transverse stretch [Eq.~\eqref{eq:D}]. The same cancellation is found at the local zeniths of L1 and V1: in each case $F_b$ is $\sim 10^{-5}$ of $|F_+|$ and $|F_\times|$, at a sky position set by that detector's location and arm orientation. Figure~\ref{fig:H1} shows $|F_b|$, $\mathcal{S}_T$, and $\mathcal{S}_b/\mathcal{S}_T$ over the celestial sphere at the same epoch ($\psi=0$) for H1, L1, and V1, with each detector's local zenith marked: regions of large tensor power coincide with suppressed breathing response at all three sites, while inclined directions restore a nonzero $F_b$ at the cost of reduced $\mathcal{S}_T$.

\subsection{Network $\mathcal{S}_b$ and inclination gain}

\begin{figure}[t]
\includegraphics[width=\columnwidth]{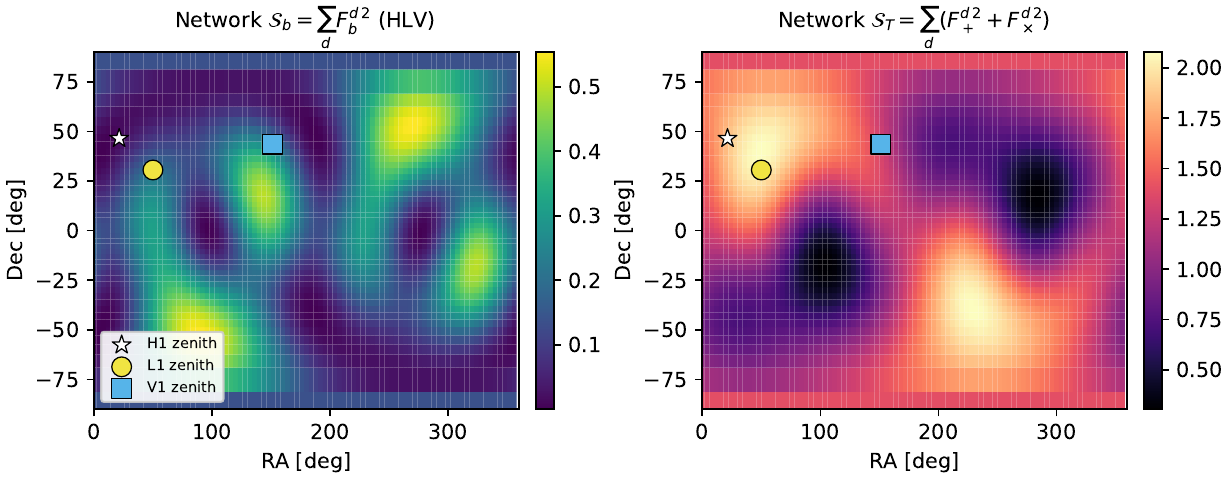}
\caption{\label{fig:HLV}
Network $\mathcal{S}_b$ and $\mathcal{S}_T$ for H1+L1+V1 at the GW150914 epoch
(PyCBC; GPS $1126259462.4$, $\psi=0$).
Left: breathing power $\mathcal{S}_b^{\mathrm{net}}=\sum_d F_b^{d\,2}$;
right: tensor power
$\mathcal{S}_T^{\mathrm{net}}=\sum_d(F_+^{d\,2}+F_\times^{d\,2})$.
Markers show the local zeniths of H1 (star), L1 (circle), and V1 (square).
Distinct arm orientations redistribute sensitivity, but the two maps peak in
different sky sectors.}
\end{figure}

For a three-site H1+L1+V1 (HLV) combination the network powers
\begin{equation}
  \mathcal{S}_b^{\mathrm{net}}=\sum_d F_b^{d\,2},\qquad
  \mathcal{S}_T^{\mathrm{net}}=\sum_d\bigl(F_+^{d\,2}+F_\times^{d\,2}\bigr)
\end{equation}
remain strongly anisotropic (Fig.~\ref{fig:HLV}): the left panel maps network $\mathcal{S}_b^{\mathrm{net}}$ and the right panel $\mathcal{S}_T^{\mathrm{net}}$, which peak in different sky sectors (local zeniths of H1, L1, and V1 are marked). Distinct arm orientations redistribute sensitivity; the preference for inclined skies remains.

\begin{figure}[t]
\includegraphics[width=\columnwidth]{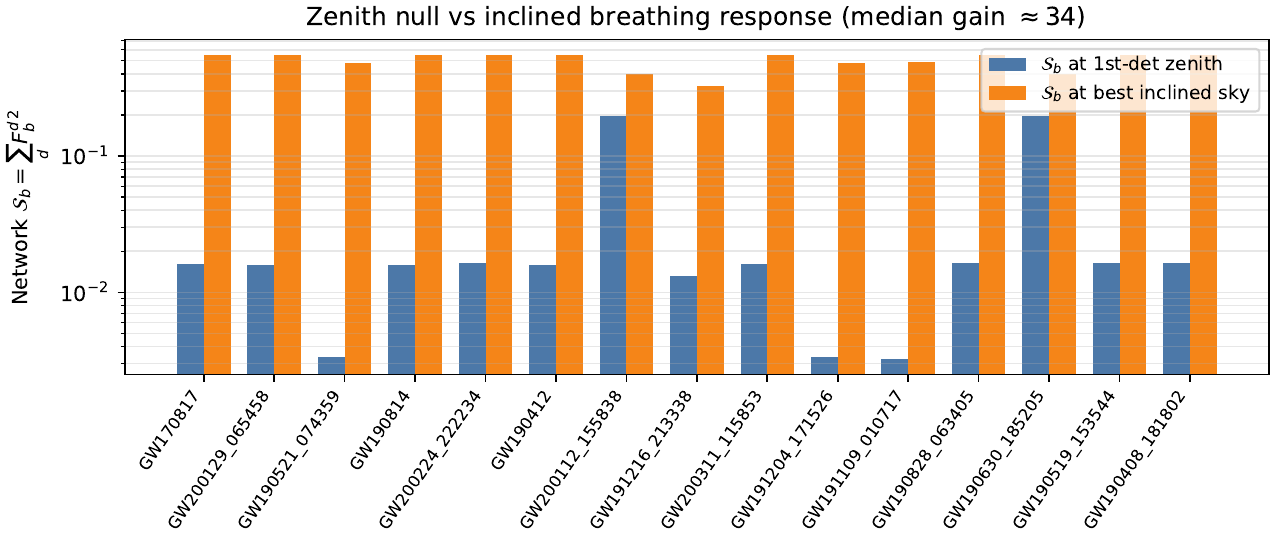}
\caption{\label{fig:gain}
Paired comparison of network $\mathcal{S}_b=\sum_d F_b^{d\,2}$ for the
fifteen loudest GWTC events with network $\mathrm{SNR}\ge 12$ (log scale):
at the local zenith of the first contributing detector (left bar of each
pair) versus at the celestial-grid sky that maximizes
$\mathcal{S}_b\,(\mathcal{S}_T/(\mathcal{S}_T+0.05))$ (right bar).
The median inclination gain over the $72$-event sample
($\mathrm{SNR}\ge 8$, $\ge 2$ sites) is $\approx 34$.}
\end{figure}

For each of $72$ GWTC events with $\mathrm{SNR}\ge 8$ and at least two LVK sites, $\mathcal{S}_b^{\mathrm{net}}$ is compared (i)~at the local zenith of the first contributing detector and (ii)~at the sky position that maximizes $\mathcal{S}_b\,(\mathcal{S}_T/(\mathcal{S}_T+0.05))$ on a celestial grid. The factor $\mathcal{S}_T/(\mathcal{S}_T+0.05)$ is a soft tensor floor: the additive $0.05$ is a small regularizer in antenna power that keeps the maximizer away from tensor-blind skies. Over this full sample the median inclination gain $\mathcal{S}_b^{\mathrm{(inclined)}}/\mathcal{S}_b^{\mathrm{(zenith)}}$ is $\approx 34$. Figure~\ref{fig:gain} shows the same zenith--inclined pairing, on a log scale, for the fifteen loudest events with network $\mathrm{SNR}\ge 12$; the bar heights vary from event to event, so the quoted median is that of the $72$-event sample.

\subsection{Catalog ranking for $\xi$ reach}
\label{sec:fom}

Beyond the $72$ events with packaged $4\,\mathrm{kHz}$ event-page strain ($32\,\mathrm{s}$ files), the full multi-detector GWTC table is ranked by the geometric proxy
\begin{equation}
  \mathrm{SNR}\,\sqrt{\mathcal{S}_b^{\,*}}
\end{equation}
for the breathing reach under $h_b=\xi h_+$, where $\mathcal{S}_b^{\,*}$ is the network breathing power at the best inclined sky (same soft tensor floor as above). For $\mathrm{SNR}\ge 8$ and at least two of $\{\mathrm{H1},\mathrm{L1},\mathrm{V1},\mathrm{K1}\}$ this cut on the public GWTC tables yields $355$ events. Table~\ref{tab:fom} reports only the twelve highest scores. Two-site HL rows share $\mathcal{S}_b^{\,*}\sim 0.48$, three-site HLV rows $\sim 0.55$; the HV event GW241011\_233834 has a lower $\mathcal{S}_b^{\,*}=0.323$, so its rank is SNR-driven. Several triggers from the fourth observing run (O4) and GWTC-4/5 (notably GW250114\_082203~\cite{LIGOScientific:2026gwtc5,LIGOScientific:2025gw250114_discovery,LIGOScientific:2025gw250114_spectroscopy}) outrank events from the third observing run (O3) by raw SNR, while GW170817 and GW200129\_065458 remain among the top geometric targets with packaged strain. This ranking is an optimistic upper bound on $\xi$ sensitivity because the true PE sky can differ from the best inclined grid point.

\begin{table}[t]
\caption{\label{tab:fom}
Top geometric ranking by $\mathrm{SNR}\sqrt{\mathcal{S}_b^{\,*}}$ among
multi-detector GWTC events with catalog network $\mathrm{SNR}\ge 8$
and at least two of $\{\mathrm{H1},\mathrm{L1},\mathrm{V1},\mathrm{K1}\}$
($355$ events under this public-catalog cut; only the twelve highest
scores are listed).
``Net.'' lists the active interferometers (H/L/V/K shorthand).
$\mathcal{S}_b^{\,*}$ is the network breathing power at the best inclined
sky (soft tensor floor defined above).
}

\begin{ruledtabular}
\begin{tabular}{lcccc}
Event & Net. & SNR & $\mathcal{S}_b^{\,*}$ & $\mathrm{SNR}\sqrt{\mathcal{S}_b^{\,*}}$  \\
\colrule
GW250114\_082203 & HL & $78.6$ & $0.481$ & $54.5$ \\
GW230814\_230901 & HL & $43.0$ & $0.479$ & $29.8$ \\
GW240920\_124024 & HL & $37.4$ & $0.480$ & $25.9$ \\
GW170817 & HLV & $33.0$ & $0.550$ & $24.5$  \\
GW231226\_101520 & HL & $34.7$ & $0.484$ & $24.1$ \\
GW241127\_061008 & HLV & $31.3$ & $0.550$ & $23.2$ \\
GW240925\_005809 & HLV & $31.2$ & $0.550$ & $23.1$ \\
GW240621\_195059 & HLV & $28.2$ & $0.549$ & $20.9$ \\
GW241011\_233834 & HV & $35.8$ & $0.323$ & $20.4$ \\
GW230627\_015337 & HL & $28.7$ & $0.482$ & $19.9$ \\
GW200129\_065458 & HLV & $26.8$ & $0.552$ & $19.9$  \\
GW240615\_113620 & HLV & $26.4$ & $0.549$ & $19.6$ \\
\end{tabular}
\end{ruledtabular}
\end{table}

\subsection{PE-weighted network $\mathcal{S}_b$}
\label{sec:pe-fb}

Catalog sky points and the best-sky ranking omit localization uncertainty. Preferred GWTC PE releases for twelve loud multi-detector events are used, and $\mathcal{S}_b^{\mathrm{net}}$ and $\mathcal{S}_T^{\mathrm{net}}$ are evaluated on each posterior sample in $(\mathrm{ra},\mathrm{dec},\psi)$ at the event coalescence time with the same LVK antenna patterns (via PyCBC). Table~\ref{tab:pe-fb} ranks the events by the PE-informed proxy $\mathrm{SNR}\sqrt{\mathcal{S}_b}$ (median $\mathcal{S}_b$). GW200129\_065458, GW150914, and GW190814 lead. GW200311\_115853 has a comparable median $\mathcal{S}_b$ ($0.301$) but a lower catalog SNR, so it sits fourth on the proxy. GW170817, despite a top geometric rank in Table~\ref{tab:fom}, has median $\mathcal{S}_b\approx 0.045$ and $P(\mathcal{S}_b>0.05)=0$, so its PE sky sits near a Michelson breathing null. Here $P(\mathcal{S}_b>0.05)$ is a reporting fraction above a fixed antenna-power cut, distinct from the ranking regularizer $0.05$ above; values $0.60$--$0.66$ (GW190521\_074359, GW191109\_010717, GW200112\_155838) mean the PE sky straddles that cut. The $\mathcal{S}_b/\mathcal{S}_T$ column is the median ratio of breathing to tensor antenna power on the same draws; several HL or HV events have $5\%$--$95\%$ spans that are a large fraction of the median (e.g., GW191109\_010717), so the PE sky spans a range of $\mathcal{S}_b$. Where Virgo participates, $|F_b|$ is often Virgo-dominated under the same PE draws (e.g., GW190814: median $|F_b^{\mathrm{V1}}|\approx 0.47$ versus $\lesssim 0.1$ at H1/L1).

\begin{table}[]
\caption{\label{tab:pe-fb}
PE-weighted network breathing metrics from $N=3000$ preferred-PE draws in
$(\mathrm{ra},\mathrm{dec},\psi)$ at coalescence, ordered by the proxy
$\mathrm{SNR}\sqrt{\mathcal{S}_b}$ using the catalog network SNR and the
median $\mathcal{S}_b$.
``Net.'' lists the active interferometers.
Quoted $\mathcal{S}_b$ and $\mathcal{S}_b/\mathcal{S}_T$ are medians with
$5\%$--$95\%$ quantiles;
$P(\mathcal{S}_b{>}0.05)$ is the fraction of draws above a reporting
antenna-power cut.
}
\begin{ruledtabular}
\begin{tabular}{lc cccc}
Event & Net. & $\mathcal{S}_b$ & $\mathcal{S}_b/\mathcal{S}_T$ &
$P(\mathcal{S}_b{>}0.05)$ & $\mathrm{SNR}\sqrt{\mathcal{S}_b}$ \\
\colrule
GW200129\_065458 & HLV & $0.310^{+0.004}_{-0.089}$ & $0.234^{+0.018}_{-0.012}$ & $1.00$ & $14.9$ \\
GW150914 & HL & $0.258^{+0.057}_{-0.127}$ & $0.322^{+0.411}_{-0.197}$ & $1.00$ & $13.2$ \\
GW190814 & HLV & $0.234^{+0.012}_{-0.056}$ & $0.119^{+0.008}_{-0.029}$ & $1.00$ & $12.3$ \\
GW200311\_115853 & HLV & $0.301^{+0.057}_{-0.062}$ & $0.281^{+0.042}_{-0.053}$ & $1.00$ & $9.8$ \\
GW191216\_213338 & HV & $0.246^{+0.051}_{-0.173}$ & $0.256^{+0.267}_{-0.192}$ & $0.97$ & $9.2$ \\
GW190521\_074359 & HL & $0.088^{+0.146}_{-0.086}$ & $0.073^{+0.287}_{-0.071}$ & $0.66$ & $7.7$ \\
GW190412 & HLV & $0.136^{+0.009}_{-0.004}$ & $0.073^{+0.005}_{-0.003}$ & $1.00$ & $7.3$ \\
GW170817 & HLV & $0.045^{+0.000}_{-0.000}$ & $0.031^{+0.000}_{-0.000}$ & $0.00$ & $7.0$ \\
GW191109\_010717 & HL & $0.154^{+0.310}_{-0.148}$ & $0.156^{+0.794}_{-0.152}$ & $0.63$ & $6.8$ \\
GW200224\_222234 & HLV & $0.112^{+0.030}_{-0.019}$ & $0.079^{+0.027}_{-0.016}$ & $1.00$ & $6.7$ \\
GW190828\_063405 & HLV & $0.131^{+0.023}_{-0.074}$ & $0.071^{+0.014}_{-0.042}$ & $0.97$ & $6.0$ \\
GW200112\_155838 & LV & $0.072^{+0.126}_{-0.061}$ & $0.083^{+0.135}_{-0.067}$ & $0.60$ & $5.3$ \\
\end{tabular}
\end{ruledtabular}
\end{table}

\subsection{Injection sensitivity grid}
\label{sec:bilby}

As a calibration of Bayesian model comparison, a non-precessing HLV IMRPhenomD~\cite{IMRPhenomD} CBC is injected into Gaussian noise colored by the design power spectral densities (PSDs), $h_b=\xi\,h_+$ is added with the LVK breathing antenna pattern $F_b$ (via PyCBC), and nested sampling is performed with bilby~\cite{Ashton2019} under GR ($\xi\equiv 0$) versus GR+$b$ ($\xi$ free on $[-1.5,1.5]$). Masses, spins, luminosity distance $d_L$, and inclination $\iota$ are fixed; coalescence time $t_c$, sky, phase, and (when free) $\xi$ are sampled. Here $\iota$ is the source inclination in the injection waveform, playing the same role as $\theta_{JN}$ in the PE-informed analyses below. The injection amplitude is scanned over $\xi_{\mathrm{inj}}\in\{0,0.5,1\}$ on an inclined sky ($\mathcal{S}_b\approx 0.18$) and near the H1 zenith ($\mathcal{S}_b\approx 0.016$), at tensor network SNR~$\sim 23$. Table~\ref{tab:inj-grid} summarizes the evidences. On the inclined sky, $\xi=1$ yields a log Bayes factor $\ln B(\mathrm{GR}{+}b/\mathrm{GR})\approx 13.2$ and a recovered posterior centered on the injection ($0.97_{-0.31}^{+0.25}$); $\xi=0.5$ is already preferred ($\ln B\approx 3.8$, median $0.45_{-0.22}^{+0.30}$), while $\xi=0$ mildly favors GR. Near zenith the same $\xi=1$ injection is inconclusive ($\ln B\sim\mathcal{O}(0)$). The three zenith posteriors all have medians $\sim 0.35$--$0.54$ and $90\%$ half-widths $\sim 0.7$, irrespective of $\xi_{\mathrm{inj}}$: once $\mathcal{S}_b$ is suppressed, the posterior on $\xi$ remains unconstrained. Geometric $\mathcal{S}_b$, rather than SNR alone, sets the reach in $\xi$.

\begin{table}[t]
\caption{\label{tab:inj-grid}
Injection calibration of GR versus GR+$b$ for an HLV IMRPhenomD CBC in
design-sensitivity Gaussian noise (tensor network SNR~$\sim 23$;
shared noise realization within each sky block; $d_L=2000\,\mathrm{Mpc}$).
Masses, spins, $d_L$, and inclination $\iota$ are fixed; coalescence time $t_c$, sky, phase, and
(when free) $\xi\in[-1.5,1.5]$ are sampled.
$\ln B$ is the log Bayes factor $\ln B(\mathrm{GR}{+}b/\mathrm{GR})$; the last column gives the
median $\xi$ with equal-tailed $90\%$ credible interval.}
\begin{ruledtabular}
\begin{tabular}{lcccc}
Sky & $\mathcal{S}_b$ & $\xi_{\mathrm{inj}}$ &
$\ln B(\mathrm{GR}{+}b/\mathrm{GR})$ & $\xi$ (med.\ $90\%$) \\
\colrule
Inclined & $0.18$ & $0$ & $-2.9$ & $-0.03_{-0.24}^{+0.24}$ \\
Inclined & $0.18$ & $0.5$ & $+3.8$ & $0.45_{-0.22}^{+0.30}$ \\
Inclined & $0.18$ & $1$ & $+13.2$ & $0.97_{-0.31}^{+0.25}$ \\
Zenith & $0.016$ & $0$ & $-1.6$ & $0.35_{-0.75}^{+0.76}$ \\
Zenith & $0.016$ & $0.5$ & $+0.6$ & $0.45_{-0.84}^{+0.73}$ \\
Zenith & $0.016$ & $1$ & $+0.1$ & $0.54_{-0.74}^{+0.62}$ \\
\end{tabular}
\end{ruledtabular}
\end{table}

\subsection{Real-event semi-joint constraints on $\xi$}
\label{sec:real-pe}
The same nested models are applied to GWTC strain for events with high PE-informed $\mathrm{SNR}\sqrt{\mathcal{S}_b}$: GW150914~\cite{Abbott_2016_GW150914}, GW200129\_065458, GW190814~\cite{Abbott_2020_GW190814}, GW191216\_213338, GW190412, GW190521\_074359, GW190828\_063405, GW191109\_010717, GW200112\_155838, GW200224\_222234, and GW200311\_115853. In this semi-joint setup, masses and spins remain fixed at preferred PE medians, while $d_L$, inclination $\theta_{JN}$, and $(\mathrm{ra},\mathrm{dec},\psi)$ are sampled within PE-informed priors, together with coalescence time $t_c$, phase, and (for GR+$b$) $\xi\in[-1.5,1.5]$. GWOSC $4\,\mathrm{kHz}$ strain is analyzed in an $8\,\mathrm{s}$ window for all events except GW190814, which uses a $16\,\mathrm{s}$ window because of its low chirp mass and correspondingly longer in-band inspiral; all runs use the IMRPhenomPv2 waveform family~\cite{IMRPhenomPv,IMRPhenomD}, matching the preferred PE releases, and PE power spectral densities where available. The injection grid above used non-precessing IMRPhenomD as a controlled calibration waveform.

\begin{table}[]
\caption{\label{tab:real-lnB}
Semi-joint real-event comparison (masses and spins fixed at preferred PE
medians; $d_L$, $\theta_{JN}$, sky, $t_c$, phase, and $\xi$ sampled within
PE-informed priors), ordered by $|\xi|_{90\%}$.
$\ln B$ is $\ln B(\mathrm{GR}{+}b/\mathrm{GR})$ with nested-sampling
uncertainty (quadrature sum of the GR and GR+$b$ evidence errors); $\xi$ is
quoted as median with equal-tailed $90\%$ interval;
$|\xi|_{90\%}$ is the one-sided $90\%$ upper bound on $|\xi|$.}
\begin{ruledtabular}
\begin{tabular}{lccc}
Event & $\ln B$ & $\xi$ (med.\ $90\%$) & $|\xi|_{90\%}$ \\
\colrule
GW150914 & $-2.19\pm0.7$ &
$0.01_{-0.15}^{+0.14}$ & $0.15$ \\
GW200129\_065458 & $-0.61\pm0.6$ &
$0.02_{-0.16}^{+0.19}$ & $0.18$ \\
GW200311\_115853 & $-2.04\pm0.8$ &
$0.06_{-0.25}^{+0.20}$ & $0.23$ \\
GW190814 & $-1.50\pm0.6$ &
$-0.11_{-0.56}^{+0.61}$ & $0.61$ \\
GW191216\_213338 & $-1.37\pm0.6$ &
$-0.04_{-0.49}^{+0.85}$ & $0.67$ \\
GW200224\_222234 & $+0.38\pm0.8$ &
$-0.16_{-0.57}^{+0.81}$ & $0.71$ \\
GW190412 & $+0.01\pm0.8$ &
$-0.26_{-0.66}^{+0.69}$ & $0.74$ \\
GW200112\_155838 & $-0.46\pm0.9$ &
$0.05_{-0.90}^{+0.98}$ & $0.93$ \\
GW190521\_074359 & $-0.62\pm0.9$ &
$0.12_{-0.92}^{+1.05}$ & $1.01$ \\
GW190828\_063405 & $-1.10\pm0.9$ &
$-0.00_{-0.98}^{+1.05}$ & $1.04$ \\
GW191109\_010717 & $-0.41\pm0.9$ &
$-0.07_{-1.08}^{+1.26}$ & $1.18$ \\
\end{tabular}
\end{ruledtabular}
\end{table}

\begin{figure}[t]
\includegraphics[width=\columnwidth]{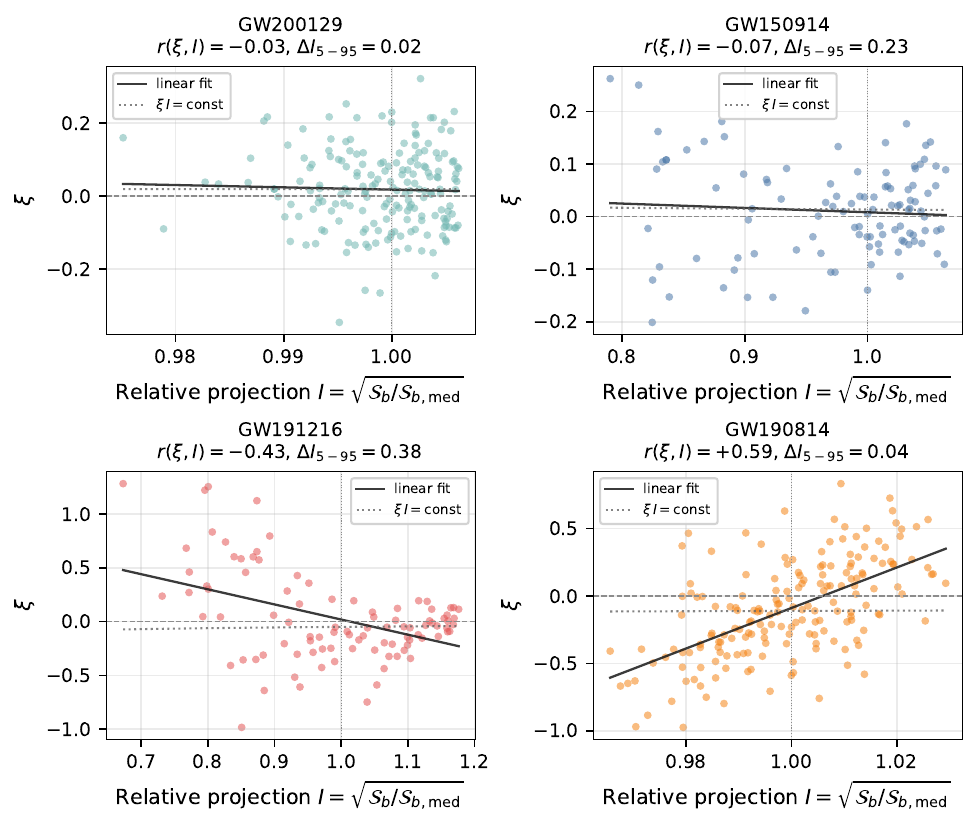}
\caption{\label{fig:xi-Sb}
Semi-joint GR+$b$ posteriors in the plane of relative breathing projection
$I=\sqrt{\mathcal{S}_b/\mathcal{S}_{b,\mathrm{med}}}$ versus $\xi$ for
GW150914, GW200129\_065458, GW190814, and GW191216\_213338.
The four events differ in sky-driven $I$ span; Table~\ref{tab:real-lnB}
lists tighter $|\xi|_{90\%}$ for other events.
Here $\mathcal{S}_b=\sum_d F_b^{d\,2}$ is recomputed from each sample's
$(\mathrm{ra},\mathrm{dec},\psi)$ at fixed GPS, so $I$ isolates sky-driven
antenna variation (vertical dotted line: $I=1$; horizontal dashed line:
$\xi=0$).
Panels are ordered by increasing Pearson $|r(\xi,I)|$; each title quotes the
coefficient and the sky-driven intensity span $\Delta I_{5{-}95}$.
Solid curves are linear fits; dotted curves show $\xi I=\mathrm{const}$
(product degeneracy).}
\end{figure}

Table~\ref{tab:real-lnB} summarizes the model comparison. None of the eleven events favors GR+$b$; two entries have $\ln B>0$ (GW200224\_222234, GW190412) but both are consistent with zero within the nested-sampling error. The $\xi$ posteriors are consistent with zero; the one-sided $90\%$ bounds reach $|\xi|_{90\%}\approx 0.15$ for GW150914, $0.18$ for GW200129\_065458, and $0.23$ for GW200311\_115853. That order differs from the PE proxy in Table~\ref{tab:pe-fb}: GW190814 ranks third there but has $|\xi|_{90\%}=0.61$, while GW200311\_115853 ranks fourth and is tighter, consistent with its higher median $\mathcal{S}_b$ ($0.301$ versus $0.234$). GW200129\_065458 has only $\ln B=-0.61\pm 0.6$, so the evidence ratio barely prefers GR, yet the $|\xi|$ interval is still narrow. Events with $|\xi|_{90\%}\gtrsim 1$ (the last four rows) still allow $|\xi|\sim\mathcal{O}(1)$. Masses and spins remain fixed, so these are semi-joint rather than full PE limits; under this model they disfavor a phenomenological breathing amplitude $|\xi|\sim\mathcal{O}(1)$ for the three tightest events. Because the breathing contribution scales as $\xi F_b$, PE sky samples change the relative projection strength $I=\sqrt{\mathcal{S}_b/\mathcal{S}_{b,\mathrm{med}}}$ and can partially trade off against $\xi$ (approximate product degeneracy $\xi I\sim\mathrm{const}$). Figure~\ref{fig:xi-Sb} shows the semi-joint GR+$b$ posteriors in the $(I,\xi)$ plane for GW150914, GW200129\_065458, GW190814, and GW191216\_213338, whose sky-driven $I$ spans differ; GW200311\_115853 is omitted because the figure is ordered by $I$ variation. $\mathcal{S}_b$ (hence $I$) is recomputed from each sample's $(\mathrm{ra},\mathrm{dec},\psi)$ at fixed GPS, so the abscissa is the sky-driven antenna variation. Panels are ordered by increasing Pearson correlation $|r(\xi,I)|$. GW200129\_065458 has a negligible $I$ span ($\Delta I_{5{-}95}\approx 0.02$, $|r|\approx 0.03$); GW191216\_213338 has the largest sky-driven $\Delta I$ ($\approx 0.38$) and an anti-correlation $r\approx -0.43$, consistent with $\xi I$ degeneracy (dotted curves). GW150914 shows a sizable $\Delta I$ but weak correlation ($|r|\approx 0.07$); GW190814 shows a positive $r\approx +0.59$ over a narrow $I$ range, unlike $\xi I=\mathrm{const}$. The reported $|\xi|_{90\%}$ limits already marginalize over this coupling. They sit alongside catalog pure-polarization and null-stream tests~\cite{Abbott2019TGR,Abbott2021TGR,Cheung2025}, which constrain more general non-tensor content. The present analysis constrains the phenomenological extension $h_b=\xi h_+$, including the scale $|\xi|\sim\mathcal{O}(1)$ taken from GQFT.

\subsection{O4 high-ranking targets}

With \S\ref{sec:real-pe} now covering eleven of the twelve top PE-weighted targets in Table~\ref{tab:pe-fb} (GW170817 excluded because its PE sky sits on the breathing null), the remaining high-ranking events in Table~\ref{tab:fom} await $\xi$ limits. O4/GWTC-4/5 triggers dominate that geometric ranking by $\mathrm{SNR}\sqrt{\mathcal{S}_b^{\,*}}$. They typically lack the older GWOSC event-page Hierarchical Data Format (HDF5) files (sampled at $4\,\mathrm{kHz}$, duration $32\,\mathrm{s}$). The semi-joint analysis uses $4\,\mathrm{kHz}$ strain in $8$--$16\,\mathrm{s}$ windows. GWOSC $4\,\mathrm{kHz}$ archive segments spanning several thousand seconds are obtained for the loudest accessible targets (GW250114\_082203, GW240920\_124024, GW230814\_230901, GW241011\_233834), and windows of $\pm 16\,\mathrm{s}$ about the trigger time are extracted. Dual-detector usable strain is available for GW250114\_082203 and GW240920\_124024 (H1+L1); GW230814\_230901 has usable L1 but a gated H1 window near the trigger in the GWOSC archive. PE-weighted $\mathcal{S}_b$ and semi-joint $\xi$ limits for these O4 events are deferred until the corresponding PE skies can be propagated through the same $\mathrm{SNR}\sqrt{\mathcal{S}_b}$ selection. Table~\ref{tab:fom} identifies them as the remaining targets.

\section{Conclusions}
\label{sec:concl}

Compact-binary detections in the LVK network are described by the tensor pair $+,\times$. Extra breathing radiation is allowed in scalar--tensor extensions, yet solar-system bounds already push that amplitude well below the tensor sector for observationally allowed couplings, so a null result is expected whenever $|\xi|\ll 1$. GQFT admits $|\xi|\sim\mathcal{O}(1)$ for $h_b=\xi h_+$~\cite{Gao2024,Gao:2025aye,XuJinWu2025}. On terrestrial interferometers that scale is further limited by geometry: directions of peak tensor response coincide with a near cancellation of the isotropic $b$ mode, and the network combination $\mathcal{S}_b$ measures how much breathing power remains.

Catalog tests of GR and null-stream searches constrain pure-polarization hypotheses or fractional non-tensor strain; they leave unconstrained a CBC template in which $b$ is tied to $h_+$. Space-based maps at $|\xi|\sim\mathcal{O}(1)$ exist, but omit Michelson zenith suppression. The present work constrains phenomenological $\xi$ with GWTC open data after the terrestrial antenna pattern is included.

Antenna patterns are taken from the PyCBC realization of the detector tensor. Geometric scans of packaged-strain events and PE posterior skies supply $\mathcal{S}_b$; events are then ordered by a best-sky proxy $\mathrm{SNR}\sqrt{\mathcal{S}_b^{\,*}}$. Nested GR versus GR+$b$ comparisons use bilby. On the injection grid, masses, spins, distance, and inclination are held fixed, while sky and $\xi$ vary. On GWOSC strain the same nested models are semi-joint: masses and spins stay at preferred PE medians, while distance, inclination, and sky vary inside PE-informed priors, so that the inferred $\xi$ already averages over projection strength.

At fixed tensor SNR the reach in $\xi$ is set by $\mathcal{S}_b$. At the H1 zenith at the GW150914 epoch, $F_b\sim 5\times 10^{-6}$ (Fig.~\ref{fig:H1}), and the packaged-strain sample shows a median $\sim 34\times$ gain in $\mathcal{S}_b$ once the sky is inclined (Fig.~\ref{fig:gain}). Optimistic ranking of the multi-detector GWTC cut ($\mathrm{SNR}\ge 8$, $\ge 2$ sites) therefore highlights loud O4 triggers (Table~\ref{tab:fom}), yet PE localization can move a geometrically attractive source onto a breathing null, as for GW170817 (Table~\ref{tab:pe-fb}). Injections at fixed tensor SNR recover $\ln B(\mathrm{GR}{+}b/\mathrm{GR})\approx 13$ ($4$) for $\xi=1$ ($0.5$) only on an inclined sky; near zenith the $\xi$ posterior is unconstrained regardless of the injected value (Table~\ref{tab:inj-grid}). Semi-joint runs on eleven events yield no preference for $\mathrm{GR}{+}b$. The PE-informed proxy in Table~\ref{tab:pe-fb} differs from the $|\xi|_{90\%}$ order: the three tightest bounds, $|\xi|_{90\%}\approx 0.15$, $0.18$, and $0.23$ (GW150914, GW200129\_065458, GW200311\_115853), lie below $|\xi|\sim\mathcal{O}(1)$ after the sky-driven factor $I$ is marginalized (Table~\ref{tab:real-lnB}, Fig.~\ref{fig:xi-Sb}), whereas weaker bounds, including GW190814, still allow that scale. High-ranking O4 triggers remain for the same pipeline once public PE skies are available.

Under this model those three events are consistent with GR and disfavor a breathing strain comparable to $h_+$. The limits sit alongside existing evidence for tensor-dominated LVK signals~\cite{Abbott2019TGR,Abbott2021TGR,Pang2020,Cheung2025,Hagihara2019} and the space-detector forecasts at $|\xi|\sim\mathcal{O}(1)$~\cite{XuJinWu2025}. Freeing masses and spins, enlarging the open-data sample, and studying waveform systematics will further constrain $\xi$.
\begin{acknowledgments}
We thank LVK collaboration members for discussions. Hong-Bo Jin is funded by the National Astronomical Observatories of the Chinese Academy of Sciences, Project No.~E4TG6601, and has been supported in part by the National Key Research and Development Program of China under Grant No.~2021YFC2203000. This material is based upon work supported by NSF's LIGO Laboratory which is a major facility fully funded by the National Science Foundation.
\end{acknowledgments}

\bibliography{ref}

\end{document}